\documentclass[aps,11pt]{article}

\usepackage{graphicx}

\newcommand{\ba} {\begin{eqnarray}}
\newcommand{\ea} {\end{eqnarray}}
\newcommand{\be} {\begin{equation}}
\newcommand{\ee} {\end{equation}}
\begin{document}

\title{\vspace*{-4cm}\hfill {\small\bf ECTP-2010-1}\\
\vspace*{4cm} Thermodynamics in the Viscous Early Universe}

\author{A.~Tawfik\thanks{drtawfik@mti.edu.eg}\\~\\
{\small Egyptian Center for Theoretical Physics (ECTP), MTI University,
 Cairo-Egypt} 
}

\date{}
\maketitle

\begin{abstract}
Assuming that the matter filling the background geometry in the Early Universe
was a free gas and no phase transitions took place, we
discuss the thermodynamics of this closed system using classical
approaches. We found that essential cosmological quantities, such as the
Hubble parameter $H$, the scaling factor $a$ and the curvature parameter $k$,
can be derived from this simple model. The results are compatible with the 
Friedmann-Robertson-Walker model and Einstein field equations. Including
finite bulk viscosity coefficient leads to important changes in the
cosmological  
quantities. Accordingly, our picture about evolution of the Universe and its
astrophysical consequences seems to 
be a subject of radical revision. We found that $k$ strongly depends on
thermodynamics of the cosmic background matter. The time scale, at which
negative curvature might take place, depends on the relation between the
matter content and the total energy. Using quantum and statistical approaches,
we introduced expressions for $H$ and the bulk viscosity coefficient. 
\end{abstract}


\section{\label{sec:1} Introduction}

We assume that the cosmological background geometry filled with matter
or radiation is characterized by well known physical laws, like equation of
states and thermodynamics. It leads to a solid physical description for the
Universe in its very early stages. Such a description, especially in the Early
Universe, is favored, because we so far have no observational evidence against
it. Other components of the cosmological geometry, like dark matter and dark energy
wouldn't matter much during these early stages. Therefore, we can disregard them. 

In this work, we introduce a toy model based on thermodynamical approaches to
describe the Early Universe. We disregard all phase transitions and assume
that the matter filling the background geometry was likely formed as free
gas. We apply the laws of thermodynamics and fundamentals of
classical physics to derive expressions for the basic cosmological quantities,
such as the Hubble parameter $H(t)$, the scaling factor $a(t)$ and the
curvature parameter $k$. We compare them with the Friedmann-Robertson-Walker
(FRW) model and Einstein field equations. 
 
In this treatment, we apply the standard cosmological model and use natural
units in order to gain global evidences supporting the FRW model, although we
disregard the relativistic and microscopic effects. The various forms of matter
and radiation are homogeneously and isotropically distributed. We use non-relativistic
arguments to give expressions for the thermodynamic quantities in the Early Universe, which 
obviously reproduce essential parts of FRW model. We assume that the 
Universe was thermal equilibrium and therefore the interaction rates exceed
the Universe expansion rate, which was slowing down with the time $t$. Also, we
assume that the expansion was adiabatic, i.e., no entropy and heat change took place. 

Finally, we take into consideration two forms of the cosmic background
matter. The first one is ideal gaseous fluid, which is characterized by lack
of interactions and constant internal energy. The second one is viscous fluid,
which is characterized by long range correlations and velocity gradient along
the scaling factor $a(t)$.

\section{\label{sec:2} Expansion Rate in Non-Viscous Cosmology}

We assume that all types of energies in the Early Universe are heat, $Q$. In
such a closed system, the total energy is conserved, i.e.,  
\ba \label{eq:dq}
dQ &=& 0 = dU +pdV,
\ea
where $U$ is internal energy, $p$ is pressure and $V$ stands for the
volume. $V$ can be approximated as a cube with sides equal to the scaling
factor $a$, i.e., $V=a^3$ or as a sphere with radius equal to $a$, i.e., $V=(4\pi/3) a^3$. In both cases,
$V\propto a^3$. Apparently, Eq.~(\ref{eq:dq}) is the first law of
thermodynamics. The expansion of the Universe causes a change in energy
density $\rho=U/V$, i.e., surely decreasing, which can be given as $d\rho=dU/V-U
dV/V^2$. In comoving coordinates, $U$ is equivalent to the mass $m$ and
consequently to the energy. From Eq.~(\ref{eq:dq}), we get 
\ba \label{eq:dro}
d\rho &=& -3 (\rho+p)\frac{da}{a}.
\ea
Dividing both sides by an infinitesimal time element $dt$ results in
\ba \label{eq:drot}
\dot \rho &=& -3 (\rho+p)\, H,
\ea
which is the equation of motion from FRW model, which strongly depends on the
thermodynamic quantities, $\rho$ and $p$, i.e., the equation of state
(EoS). One dot means first derivative with respect to the time
$t$. $H$ is the Hubble parameter which relates velocities with distances;
$H=\dot a/a$. 

The radiation-dominated phase is characterized by $p=\rho/3$ and
therefore Eq.~(\ref{eq:drot}) leads to $\rho\propto a^{-4}$. In the 
matter-dominated phase, $p<<\rho$ and therefore $\rho\propto a^{-3}$, i.e.,
$\rho\propto V^{-1}$. The energy density $\rho$ is a function of temperature
$T$. Then, we can re-phrase the proportionality in the radiation-dominated
phase as $\rho\propto T V^{-1}$. 

Neglecting both cosmological constant $\Lambda$
and curvature parameter $k$, we simply get that $H^2\propto \rho$. Then, the scaling
factor in the radiation-dominanted phase $a\propto t^{1/2}$ and in
matter-dominanted phase $a\propto t^{2/3}$. The results are depicted in
Fig.~\ref{fig:1}.

\section{\label{sec:3} Expansion Rate in Bulk Viscous Cosmology}

Let us assume that one particle of mass $m$ is located at a distance $a$ from
some point in the Universe. Such a particle will have, in the radial
direction, kinetic energy $m\dot a^2/2$. In the opposite direction, it is
affected by a gravitational force due to its mass $m$ and the mass inside the
sphere $M=(4\pi/3)a^3 \rho$. Then the particles's {\it gravitational}
potential energy is $-GMm/a$, where $G$ is the Newtonian gravitational
constant. The total energy is
\ba \label{eq:tEnr1}
E&=&\frac{1}{2}m\dot a^2 - G\frac{Mm}{a},
\ea
which can be re-written as
\ba \label{eq:adot1}
\dot a^2 + k &=& \frac{8\pi}{3} G \rho a^2,
\ea
Last equation is, exactly, the Friedman's first equation with curvature
parameter $k=-2E/m$, which apparently refers to negative curvatures associated with
various geometrical forms depending on both total energy $E$ and mass $m$. In the 
Friedman's solution, $k$ can be vanishing or $+1$ or $-1$, referring to flat
or positively or negatively curved Universe, respectively~\cite{dverno}. Our toy model
agrees well with the negatively curved Friedman's solution, especially
when the particle mass $m$ equals two times the total energy, i.e., $m=2E$.

According to recent heavy-ion collision experiments~\cite{rhic1} and lattice QCD
simulations~\cite{nakam}, the matter under extreme conditions (very high temperature
and/or pressure) seems not to be, as we used to assume, an {\it ideal} free
gas. It is likely fluid, i.e., strongly correlated matter with finite heat
conductivity and viscosity coefficients (bulk and shear)~\cite{finiteta1}. Therefore, it is
in demand to apply this assumption on the background geometry in Early Universe. The cosmic background should not necessarily be filled with an ideal free gas. In previous
works~\cite{taw1,taw2,taw3}, we introduced models, in which we included finite
viscosity coefficient. The analytical solution of such models is a non-trivial
one~\cite{taw1,taw2,taw3}. In the present work, we try to approach the viscous
cosmology using simple models, in which we just use classical approaches. As
we have seen, the classical approaches work perfectly in the non-viscous
fluid. It is in order now to check the influence of viscous fluid on the
cosmological evolution. The simplicity of these approaches doesn't sharpen the
validity of their results. Surely, it helps to come up with ideas on reality
of the viscous cosmology. \\ 

We now assume that the test particle is positioned in a viscous surrounding. 
Then the total energy, Eq.~(\ref{eq:tEnr1}), gets an additional contribution from
the viscosity work, which apparently resists the Universe expansion,
\ba \label{eq:tEnr2}
E&=&\frac{1}{2}m\dot a^2 - G\frac{Mm}{a} - \eta a^3 \frac{\ddot a}{a},
\ea
where $\eta$ is the bulk viscosity coefficient. We assume that the expansion
of the Universe is isotropic, i.e., symmetric in all directions. Consequently,
the shear viscosity coefficient likely vanishes. Comparing
Eq.~(\ref{eq:tEnr2}) with the Friedmann's solution leads to another expression
for the curvature parameter,  
\ba \label{eq:k1}
k &=& -\frac{2E}{m}-\frac{2 \eta}{m}\frac{\ddot a}{\dot a} a^3.
\ea
Comparing Eq.~(\ref{eq:k1}) with the three values of $k$ given in the FRW model ($k={+1,0,-1}$)~\cite{dverno}, results in three expressions for the expansion rate $\dot a$ in the bulk viscous cosmology. \\

When $k=+1$, then the expansion rate or velocity reads
\ba \label{eq:adot2}
\dot a &=& \left(\frac{2E+m}{\eta}\right)\frac{1}{a^2},
\ea
It is positive everywhere and inversely proportional to $a^2$. It doesn't
depend on the comoving time $t$, directly. Apparently, its $t$-dependence is embedded in
the $t$-dependency of $E$, $m$ and $\eta$. The scaling factor itself is
\ba \label{eq:aoft2}
a(t) &=& \left(3\frac{2E+m}{\eta}\right)^{1/3} t^{1/3}.
\ea
In Fig.~\ref{fig:1}, we compare this result with the non-viscous fluid as given
in section~\ref{sec:2}, i.e., $a(t)\propto t^{1/2}$ for radiation-dominated
phase and $a(t)\propto t^{2/3}$ for matter-dominated phase. For simplicity, we
assume that all proportionality coefficients are equal. It is clear that the
scaling factor in the viscous cosmology is the slowest one. This would refer
to the fact that the viscosity likely resists the Universe
expansion. Increasing $\eta$ shrinks or shortens $a(t)$,
Eq.~(\ref{eq:aoft2}). At very small $t$, the expansion of the bulk viscous
Universe is much rapid than the other two cases (non-viscous).  

From Eqs.~({\ref{eq:adot2}) and (\ref{eq:aoft2}), the Hubble parameter reads 
\ba \label{eq:H1}
H(t) &=& \frac{1}{3t}.
\ea
Apparently, $H$ doesn't depend on any of the thermodynamic quantities. It is
always positive and decays with increasing $t$. \\

\begin{figure}
\includegraphics[width=8cm,angle=-90]{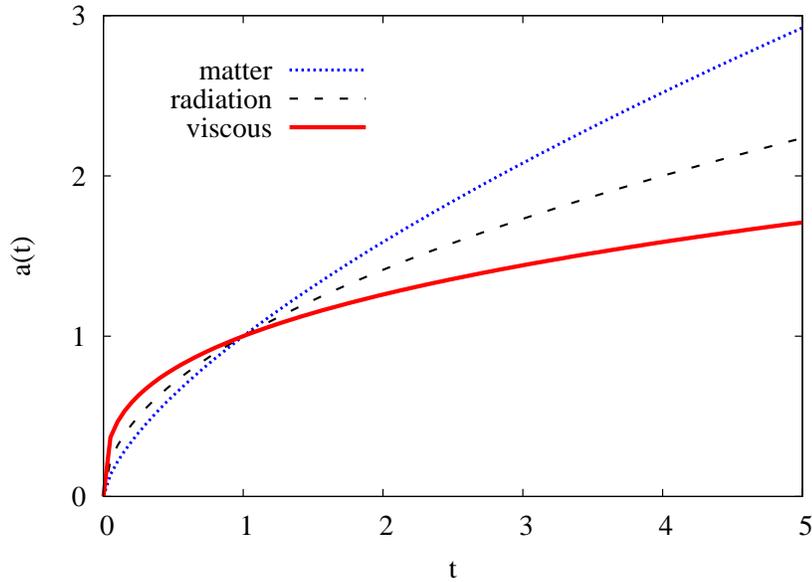}
\caption{\footnotesize Scaling factor $a$ as a function of comoving time $t$
  is depicted for viscous fluid (solid), radiation-dominanted (long dashed) and
  matter-dominanted (dotted) phases.} 
\label{fig:1}
\end{figure}

When $k=-1$, the expansion rate or velocity takes the form
\ba \label{eq:adotn}
\dot a &=& \frac{2E-m}{\eta}\frac{1}{a^2}.
\ea
It is only positive, i.e., the Universe is only expanding as long as
$m<2E$. Otherwise, the expansion rate or velocity
decreases. Eq.~(\ref{eq:adotn}) sets the limit of the Universe
contraction. This limit is reached, when the mass $m$ exceeds twice the total
energy $E$. In Early Universe, $E>>m$ and, consequently,
the Universe started explosively, although $k$ could have a negative value. Much later, $E$ decreases
according to this expansion and the matter production, meanwhile the mass $m$ gains more
and more contributions. At a certain point, the expansion rate turns to the 
backward direction. It is necessarily to mention here that, this toy model
takes into account the visible energy and matter components only. The
invisible components are not included in it.  

The scaling factor also depends on the total energy $E$ and mass $m$,
\ba \label{eq:an}
 a(t) &=& \left(3 \frac{2E-m}{\eta}\right)^{1/3} t^{1/3}.
\ea
$a$ is positive as long as $m<2E$. Otherwise it switches to negative
values. Its time dependence, $a\propto t^{1/3}$, looks like the previous case
at $k=+1$, Eq.~(\ref{eq:aoft2}).  

From Eqs.~({\ref{eq:adotn}) and (\ref{eq:an}), the Hubble parameter reads
\ba \label{eq:H2}
H(t) &=& \frac{1}{3t}.
\ea
As in the previous case, $k=+1$, $H$ is always positive and doesn't depend on
any of the thermodynamic quantities. \\

When $k=0$, the expansion rate or velocity will be
\ba \label{eq:aoft0}
\dot a &=& \frac{1}{2} \frac{E}{\eta}\frac{1}{a^2}.
\ea
In flat Universe, $\dot a$ does not depend on $m$. It increases with
increasing the total energy $E$ and decreasing the viscosity coefficient
$\eta$. Also, the scaling rate, 
\ba \label{eq:a0}
 a &=& \left(\frac{3}{2} \frac{E}{\eta}\right)^{1/3} t^{1/3},
\ea
depends on $E$ and $\eta$, only. It doesn't depend on the mass $m$, i.e., the mass production doesn't affect the scale factor or the expansion. From Eqs.~(\ref{eq:aoft0}) and (\ref{eq:a0})
\ba \label{eq:H3}
H(t) &=& \frac{1}{3\, t}.
\ea 

In Fig.~\ref{fig:2}, we depict $H(t)$ from this model and compare it with the
two cases when the background matter is a non-viscous gaseous fluid. The
latter is likely dominated by radiation or matter, where $H=1/2t$ and
$H=2/3t$, respectively. We notice that $H$ in the viscous cosmology is faster
than $H$ in the non-viscous cosmology. Therefore, we conclude that the bulk viscosity causes
slowing down the Universe expansion. \\

\begin{figure}
\includegraphics[width=8cm,angle=-90]{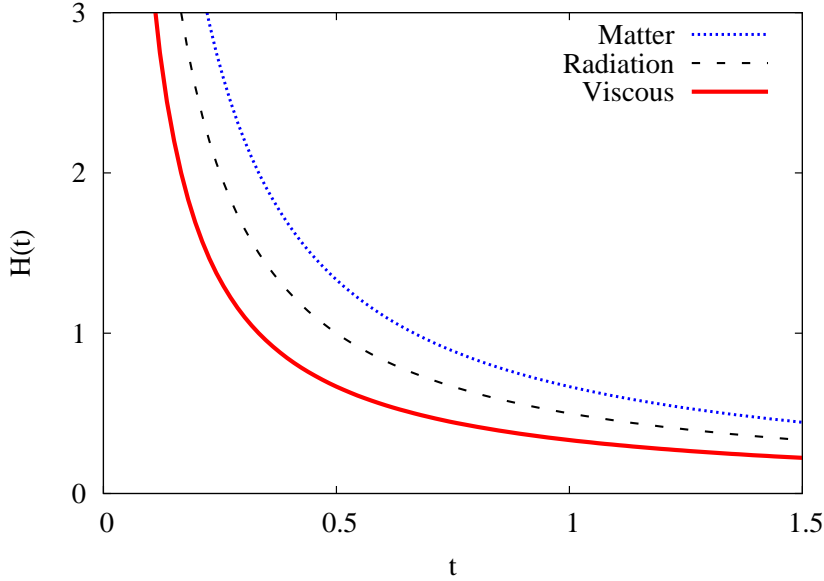}
\caption{\footnotesize The Hubble parameter $H$ as a function of comoving time
  $t$ is depicted for viscous fluid (solid) and ideal gas. The radiation- and
  matter-dominanted ideal gasses are drawn as long-dashed and dotted lines,
  respectively.} 
\label{fig:2}
\end{figure}

At $k=0$, the total energy is likely dominated by the viscosity work
\ba
E &=& -\eta a^3 \frac{\ddot a}{\dot a}.
\ea
Plugging this equation into the scaling factor, Eq.~(\ref{eq:a0}), leads to deceleration 
\ba
\ddot a &=& -\frac{2}{3} \frac{\dot a}{t}.
\ea 
Two dots refer to second derivative with respect to time $t$. \\

So far, we conclude that filling the cosmic background geometry with bulk
viscous fluid strongly affects (moderates) the expansion rate of the
Universe. The evolution of the scaling factor $a$ is damped, when $\eta$
increases. The curvature parameter $k$, which appears as a constant in FRW
model and Einstein equations~\cite{dverno}, depends on the total energy $E$,
the particle 
mass $m$ and the viscosity coefficient $\eta$. In the positively curved
Universe, $a$ increases with increasing $E$ and $m$. In the negatively
curved Universe, $a$ increases only as long as $2E>m$. Otherwise, it
decreases causing Universe contraction. We conclude that the scaling factor in
the flat Universe depends on $E/\eta$, but not on the mass constent, $m$. \\

\section{\label{sec:3b} Energy Density in Bulk Viscous Cosmology}

Including the work of bulk viscosity into the first law of thermodynamics, Eq.~(\ref{eq:dq}), results in
\ba \label{eg:du-eta}
dU &=& -\left(p\, dV + \eta\, a^3\, \frac{\ddot a}{\dot a}\, dV\right). 
\ea
Apparently, the evolution of energy density depends on the Hubble parameter $H$ and  the viscosity coefficient $\eta$,
\ba \label{eq:rhovis1}
\dot \rho &=& -\left(3(p+\rho) + 3\, \eta\, a^3\, \frac{d \dot a}{d a} \right)\, H.
\ea
Comparing this evolution equation with the one in the Eckart~\cite{Eck40}
relativistic cosmic fluid leads to a direct estimation for the bulk viscous
stress $\Pi$. The conservation of total energy density requires that the bulk
viscous stress equals to the work of bulk viscosity, i.e., $\Pi=\eta a^3 d\dot
a/d a$.  

In the radiation-dominanted phase, EoS reads $p=\rho/3$ and
Eq.~(\ref{eq:rhovis1}) can be solved in comoving time $t$ by utilizing our
previous result on $H(t)$, Eq.~(\ref{eq:H3}) for instance, 
\ba \label{eg:rho-eta3}
\rho &=& -\ln t^{(12p-\eta/3)}.
\ea
On one hand, it implies that $\rho$ diverges at $t=0$. On the other hand, the
viscosity coefficient $\eta$ seems to moderate the evolution of the total
energy density. Obviously, this result strongly depends on
EoS~\cite{taw1,taw2,taw3}, which is different in the different phases of Early
Universe, i.e. differs with $t$. 

Eq.~(\ref{eq:rhovis1}) is consistent with the second law of thermodynamics at
non-negative entropy production, $S^i_{;i}=\Pi^2/\eta T \ge0$. In this model,
$S^i_{;i} \propto V H/T$. 
The Friedmann's second solution in flat Universe, whose background geometry is
filled with a non-viscous fluid, 
\ba \label{eq:2nd1}
\frac{\ddot a}{a} &=& - \frac{4 \pi}{3} \, G \, (p+\rho).
\ea
seems to follow the second law of thermodynamics. To show this, let us take
the time derivative of last expression. Then $d (\rho\, a^3)\equiv - a^2
(p+\rho) d a$. Depending on EoS, for instance in de Sitter Universe, last
equivalence can be re-written as 
\ba
d (\rho\, a^3) &=& - 3 a^2 p\, da.
\ea
It is nothing but the second law of thermodynamics ($dU=-pdV+TdS$) of an adiabatic system, i.e., the expansion is thermally reversible and obviously doesn't affect the entropy content, $dS=0$.

\section{\label{sec:4} Hubble Parameter in Quantum Cosmology}

Let us suppose that $N$ particles are adhered to a cubic or spherical volume,
i.e., $V\propto a^3$. The particles are distributed according an occupation
function, which depends on their quantum numbers and correlations. According
to the standard cosmological model, they are allowed to expand in a
homogeneous and isotropic way. We suppose that particles have no
interactions. Then, the energy of a single particle $E=(p^2+m^2)^{1/2}$, where
momentum $\vec{p}=\frac{h}{a}(n_1\hat{x}+n_2\hat{y}+n_3\hat{z})$. In natural
units, $h=2\pi$. The state density in momentum space
$a^3/h^3=V/(2\pi)^3$. From the integral of particle density in phase space, we
get the particle density in ordinary space and, therefore, 
\ba
a^3 &=& N \left.\frac{2\pi^2}{g} \right/\left.\int_0^{\infty} \frac{p^2
  dp}{e^{\frac{E-\mu}{T}}\pm 1} \right.,
\ea
where $\mu$ is the chemical potential and $g$ is the degeneracy factor. 
Taking the time derivative (equivalent to $1/T$) and dividing both sides by the scaling factor $a$ results in 
\ba 
H(T,t) &=& \frac{1}{n\,T^2} \left\{\frac{1}{6} \frac{g}{2\pi^2}
\int_0^{\infty} \frac{\mu-E}{1+ \cosh(\frac{E-\mu}{T})} \;p^2\,dp\right\}
\frac{dT}{dt}, \label{eq:Hquant2} 
\ea
where $n$ is the particle number density. $n$ depends on the intensive state
variables, $T$, $p$ and $\mu$. It implies that $H$ depends on $1/n T^2$ and
the time derivative, $dT/dt$, besides the integral, which can be calculated,
numerically, in dependence on $T$ and $\mu$. It is obvious that the expansion
of the Universe is driven by generating new states. \\

When assuming that the background geometry is filled with a relativistic
Boltzmann's gas, then the equilibrium pressure and energy density at vanishing
viscosity are given as 
\ba \label{eq:pE1}
p(m,T) &\approx&nT, \nonumber \\
\epsilon(m,T) &\approx& n \left(\frac{3T}{m}+ \frac{K_1(m/T)}{K_2(m/T)}
\right)\, m, 
\ea
where $K_i$ is the $i$-th order modified Bessel function. At equilibrium, 
the entropy is maximum. At vanishing chemical potential, the Hubble
parameter in Eq.~(\ref{eq:Hquant2}) reads
{\footnotesize\bf
\ba \label{eq:Hquant3}
H &=& -\frac{g}{2\pi^2}\frac{1}{6n} \left[p E + 4T {\cal M}
 \tanh^{-1}\left(\frac{2 p T}{E {\cal M}}\right) -
 \left(m^2+8T^2\right)\ln(2p+2E) 
 \right] \frac{dT}{dt}, \nonumber
\ea }
where ${\cal M}=(m^2+8T^2)^{1/2}$. In the relativistic limit, i.e.,
 $m\rightarrow 0$,  
\ba \label{eq:Hquant4}
H &=& -\frac{g}{2\pi^2}\frac{1}{6n} \left[p^2 + 8 T^2\left(\sqrt{2}
 \tanh^{-1}\frac{1}{\sqrt{2}} - \ln(4p)\right) 
 \right] \frac{dT}{dt}.
\ea 
We conclude that $H$ in non-viscous quantum cosmology $H$ depends on the
 intensive state quantity $T$, its decay with the time $t$, state density in
 momentum space and both of momentum and number of occupied
 states~\cite{taw-new}.

\section{\label{sec:5} Bulk Viscosity in Quantum System}

In this section, we give estimates for the bulk viscous coefficient in both of
quark-gluon plasma and hadrons, which can be inserted in Eq.~(\ref{eg:du-eta})
to calculate the expressions given in
Eqs.~(\ref{eq:rhovis1})~and~(\ref{eg:rho-eta3}).  

In the relativistic Boltzmann limit, the bulk viscosity~\cite{ilg1, deGroot}
reads 
\ba \label{eq:etaa1}
\eta\left(\frac{m}{T}\right)  &=& \frac{m^2}{96 \pi^2 \sigma} \;
\frac{\left\{K_2\left(\frac{m}{T}\right)\left[(5-3\gamma)\hat{h}-3\gamma
    \right] \right\}^2} {2T\,
  K_2\left(\frac{2m}{T}\right)+m\,K_3\left(\frac{2m}{T}\right)}, 
\ea
which implies that $\eta$ doesn't depend on the extensive state quantity $n$,
which gives the number of occupied states in momentum space. $\eta$ depends on
the ratio of heat capacities and enthalpy per particle, which are given by the
auxiliary functions $\gamma/(\gamma-1)=5\hat{h}-\hat{h}^2+T^2/m^2$ and
$\hat{h}=T[K_3(m/T)/K_2(m/T)]/m$, respectively. Also, it depends on mass $m$,
temperature $T$ and the cross section $\sigma$. The latter has been given in
Ref.~\cite{hisc1} and is assumed to be constant for all states or particles.  

At $m=5$ and constant $\sigma$, Eq.~(\ref{eq:etaa1}) is drawn in
Fig.~\ref{fig:4}. It is clear that $\eta(m/T)$ has two singularities, one at
$T=0$ and another one at $m/2$. It has a minimum value, at a temperature
slightly below $m/2$. At much higher temperatures, $\eta(m/T)$ increases
linearly with increasing $T$. In the high-$T$ region, Eq.~(\ref{eq:etaa1}) is
likely no longer valid. \\

\begin{figure}
\begin{center}
\includegraphics[width=8cm]{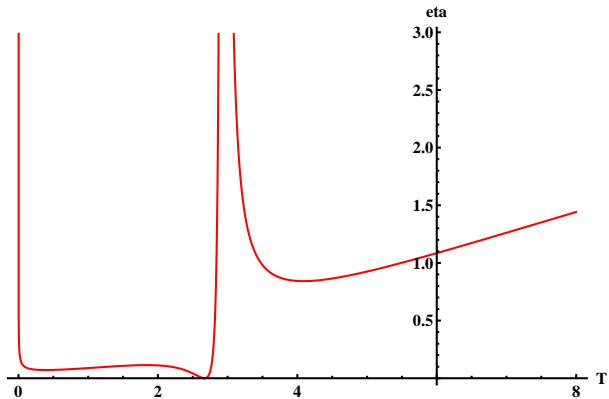}
\caption{\footnotesize At constant particle mass $m$ and cross section
  $\sigma$, $\eta$ is drawn againest $T$. There are two singularities at $T=0$
  and at $m/2$. $\eta(m/T)$ is minimum at temperatures slightly below $m/2$. } 
\label{fig:4}
\end{center} 
\end{figure}

In the Hagedorn picture, the particles, at very high energies, can be treated
non-relativistically. In this limit, we express the partition function in mass
spectrum $\rho(m)$. 
\ba
\ln Z &=& \frac{g}{2\pi^2}\,V\,T^3 \sum_{n=1}^{\infty} \rho_n(m) \frac{1}{n^2}
\left(\frac{m}{T}\right)^2 K_2\left(\frac{n\,m}{T}\right). 
\ea 
The mass spectrum $\rho(m) dm$ give the number  of states between $m$ and
$m+dm$. The non-relativistic part of the energy-momentum tensor is~\cite{ostr} 
\ba
T^{i j} &=&\frac{\tau}{(2\pi)^3} \left(\frac{i_k}{T}\right)_{,l}\int
\frac{p^{i}\, p^{j}\, p^{k}\, p^{l}}{m E} e^{-E/T} \,\rho(m)\, dm\; d^3p, 
\ea
where $\tau$ is the relaxation time. 
Taking into account the asymptotic behavior allows us to derive the bulk
viscous term in $T^{i j}$, 
\ba \label{eq:etacs}
\eta &=& \frac{5}{3}{\cal A} \frac{\tau T^{5/2}}{(2\pi)^{3/2}}
\ln\left(c_s^{-2}\right), 
\ea
where ${\cal A}$ is constant and $c_s=(\partial p/\partial \epsilon)^{1/2}$ is
the speed of sound, which characterizes the propagating of signals in the
cosmological background matter of the Early Universe. In the relativistic
limit, the partial derivatives in $c_s$ are taken, adiabatically, i.e., at
constant heat (or energy as we assumed in this model).  

$\ln(c_s^{-2})$ in Eq,~(\ref{eq:etacs}) can roughly be estimated, when we
approximate the thermodynamic quantities $p$ and $\epsilon$,
Eq.~(\ref{eq:pE1}). We assume that $p$ and $\epsilon$ are not changing with
the bulk viscous coefficient $\eta$, then 
\ba
c_s^{-2}\left(\frac{m}{T}\right) &=&  
\frac{\left(\frac{m}{T}\right)^2}{2 K_2^2 \left(\frac{m}{T}\right)} 
\left\{ K_0^2 \left(\frac{m}{T}\right) - 2 
\left(\frac{m}{T}\right)^{-1} K_0^2 \left(\frac{m}{T}\right) 
K_1^2 \left(\frac{m}{T}\right) - \right. \nonumber \\
& & \left. 2 \left[1+4\left(\frac{m}{T}\right)^{-2}\right] 
K_1^2\left(\frac{m}{T}\right) + \left[1+6\left(\frac{m}{T}\right)^{-2}\right] 
K_2^2\left(\frac{m}{T}\right) \right\}. \nonumber 
\ea
Using the dimensionless ratio $m/T$, last equation can be calculated,
numerically. This is illustrated in Fig.~\ref{fig:3}. The asymptotic value,
$\ln(3)$, is fulfilled at high $T$. When  $T\rightarrow m$, the function drops
to a minimum value. It diverges as long as $T<m$. The inverse of the
relaxation time gives the drag coefficient of the background matter. In
relativistic limit, we can model the relaxation time. For strongly coupled
${\cal N}=4$ SYM~\cite{mald}, $\tau=f(1/T)$, 
\ba\label{eq:tau1}
\tau &=& \frac{2-\ln(2)}{2\pi T}.
\ea

\begin{figure}
\begin{center}
\includegraphics[width=8cm]{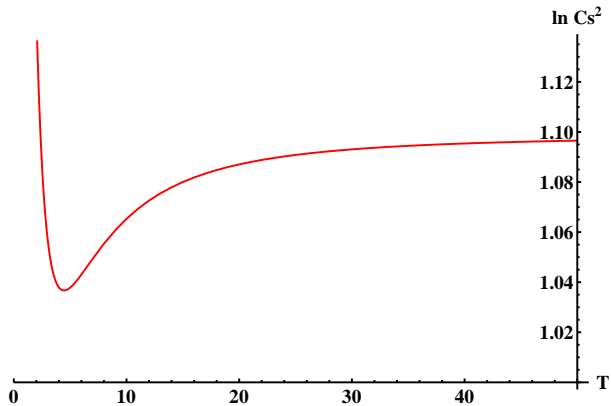}
\caption{\footnotesize At a constant particle mass, $\ln\left(c_s^{-2}\right)$
  is depicted as a function of $T$. Decreasing $T$ lowers $\ln(c_s^{-2})$
  below its asymptotic limit, $\ln(3)$. When $T\rightarrow m$,
  $\ln\left(c_s^{-2}(m/T)\right)$ drops to a minimum value. } 
\label{fig:3}
\end{center} 
\end{figure}

To keep fitting scope of present work, we leave for a future
work~\cite{taw-new}, the numerical estimates of Hubble paramter in the quantum
cosmology with (section~\ref{sec:4}) and without (section~\ref{sec:5}) bulk
viscousity.

\section{\label{sec:6} Conclusions}

We have shown that the Universe, which is characterized by the FRW model,
apparently obeys the laws of thermodynamics. We used classical assumptions in
order to derive the essential cosmological parameters, Hubble parameter $H$,
scaling factor $a$ and curvature constant $k$. In doing so, we have assumed
that the background matter is filled with an ideal thermal gas. Such a matter
is  homogeneously and isotropically distributed inside the available
cosmological geometry. For simplicity, we assume that no interactions or phase
transitions took place.   

The first gaol of this work is to study the effects of including finite bulk
viscosity on the Early Universe using classical approaches. We started from
the same assumptions as we did with the ideal thermal gas. We found
considerable changes in all cosmological parameters. Comparing our results
with the FRW model results in that the time-dependence of Hubble parameter and
scaling factor is slower than that of taking the background matter as an ideal
thermal gas.  

Also, we have found a strong dependence of $k$ on the thermodynamic
quantities, like total energy $E$ and bulk viscosity $\eta$. The relation
between $E$ and the mass $m$ determines the time scale, at which the negative
curvature sets on. The total energy in the flat Universe is characterized by a
dominant work of the bulk viscosity. When this takes place, the Universe
decelerates. Otherwise, the expansion is positive. The expansion rate is
directly proportional to $E$  and $\eta$. As for thermodynamics of the Early
Universe, we have found that the time evolution is affected by the bulk
viscosity coefficient. Should this model be considered acceptable, essential
modifications in the astrophysical observations are expected.   

The second goal of this work is to check the cosmological parameters in a
quantum system. We started from basic assumptions of quantum mechanics and
statistical physics. We expressed the Hubble parameter (and scaling factor in
a straightforward way) in dependence on the infinitesimal changes in both
phase and momentum spaces. We found that the Hubble parameter depends on $1/T$
and the time derivative of $T$. Based on this toy model, it is clear that the
expansion of the Universe is derived by the generation of new states. In the
relativistic limit, Hubble parameter depends on momentum space and the number
of occupied state, besides the decay of $T$.   

Finally, we have studied the bulk viscosity in low-$T$ and high-$T$
regimes. For the first regime, the bulk viscosity diverges at vanishing $T$
and at $T\approx m/2$, where $m$ is mass. For the high-$T$ regime, the bulk
viscosity decreases with increasing $T$. Its asymptotic value is reached, when
the speed of sound approaches its asymptotic limit. At vanishing $T$, the
speed of sound diverges. 

\section*{Acknowledgments}
This work is based on an invited talk given at the ``Second IAGA-Symposium'',
which has been held in Cairo-Egypt from 4th till 8th January 2010. I like to
thank the organizers, especially Professor A. A. Hady for the kind invitation.


\begin{thebibliography}{00}

\bibitem{dverno} R. D'Inverno, {\it Introducing Einstein's Relativity}, Oxford University Press Inc., New York (1998). 

\bibitem{rhic1} M.~Gyulassy and L.~McLerran, Nucl. Phys. A {\bf 750} 30 (2005).  

\bibitem{nakam} A.~Nakamura, S.~Sakai, Phys. Rev. Lett. {\bf 94}, 072305 (2005); S.~Sakai,, A.~Nakamura, PoS LAT2005: {\bf 186}, (2006); A. Nakamura, S.~Sakai, Nucl. Phys. A {\bf 774}, 775, (2006).

\bibitem{finiteta1} D.~Kharzeev and K. Tuchin, JHEP 0809, 093 (2008); F.~Karsch, D.~Kharzeev, K.~Tuchin, Phys. Lett. B {\bf 663}, 217, (2008) 

\bibitem{taw1} A. Tawfik, M. Wahba, H. Mansour and T. Harko, 
\newblock In Press, e-Print: arXiv:1001.2814 [gr-qc]. 

\bibitem{taw2} A. Tawfik, H. Mansour and M. Wahba, 
\newblock Talk given at 12th Marcel Grossmann Meeting on �General Relativity�, 12-18 July 2009, Paris-France, e-Print: arXiv:0912.0115 [gr-qc].

\bibitem{taw3} A. Tawfik, T. Harko, H. Mansour and M. Wahba, 
\newblock Talk at the 7th Int. Conference on �Modern Problems of Nuclear Physics�, 22-25 Sep. 2009, Tashkent-Uzbekistan, e-Print: arXiv:0911.4105 [gr-qc].

\bibitem{Eck40} C. Eckart, 
\newblock Phys. Rev. {\bf 58},  919 (1940).

\bibitem{taw-new} A. Tawfik, in progress

\bibitem{ilg1} P.~Ilg and H.~C.~Oettinger,
\newblock Phys. Rev. D {\bf 61}, 023510 (1999).

\bibitem{deGroot} S.~R.~de~Groot and W.~van~Leeuwen and C.~G.~van~Weert, 
\newblock {\it Relativistic Kinetic Theory}, North-Holland, Amsterdam, (1980).

\bibitem{hisc1} W.~Hiscock and J. Salmonson,
\newblock Phys. Rev. D {\bf 43}, 3249 (1991).

\bibitem{ostr} M.~Ostrowski, Acta Phys. Polon. B {\bf 10}, 875 (1979)

\bibitem{mald} J. M. Maldacena, Adv. Theor. Math. Phys. {\bf 2}, 231 (1998), Int. J. Theor. Phys. {\bf 38} 1113 (1999).



\end{thebibliography}
\end{document}